\documentclass{article}

\usepackage{arxiv}
\usepackage{tikz, bm, latexsym}
\usepackage{setspace, array, xcolor, enumitem}
\usepackage{amsmath, amssymb, natbib} 
\usepackage{graphicx, booktabs}
\usepackage[english]{babel}		
\usepackage{url, hyperref, placeins, rotating, parskip}
\usepackage{float}
\usepackage{natbib}

\usepackage[flushleft]{threeparttable}

\newcommand\blfootnote[1]{%
  \begingroup
  \renewcommand\thefootnote{}\footnote{#1}%
  \addtocounter{footnote}{-1}%
  \endgroup
}
\newcommand{\e}[1]{\eqref{#1}}
\newcommand{\bs}[1]{\boldsymbol{#1}}

\newtheorem{lemma}{Lemma}
\newtheorem{theorem}{Theorem}

\begin{document}

\title{Differential test functioning via robust scaling} 

\date{}

\author{Peter F. Halpin\\ 
	   School of Education\\
	   University of North Carolina at Chapel Hill \\ 
	   Chapel Hill, NC 27514 \\
	   \texttt{peter.halpin@unc.edu} 	   }

\maketitle
\begin{abstract}
In the item response theory (IRT) literature, differential test functioning (DTF) has been conceptualized in terms of how the test response function differs over groups of respondents. This paper presents an alternative approach to DTF that focusses on how the distribution of the latent trait differs over groups, which is referred to as impact. It is proposed to evaluate DTF by comparing two estimates of impact, one that naively aggregates over all test items and a robust alternative that down-weights items that exhibit differential item functioning (DIF). Taking this approach, this paper makes the following three contributions.  First it is shown that the difference between the naive and robust estimands provides a convenient effect size for quantifying the extent to which DIF affects conclusions about impact (as opposed to test scores). Second it is shown how to construct a robust estimator that yields consistent estimates of impact whenever fewer than 1/2 of items exhibit DIF. Third, a relatively general purpose Wald test of the difference between two estimates of impact is developed. Using simulations and an empirical example from physics education, it is shown how the proposed effect size and test statistic perform using the proposed robust estimator of impact, as well as estimators that arise from conventional item-by-item tests of DIF. 

\blfootnote{\\ This material is based upon work supported by the National Science Foundation under Grant No. 2400864. \\ \\

}

\end{abstract}

\newpage

\section{Introduction} 

Conventional approaches to differential test functioning (DTF) focus on how test scores differ over groups of respondents. This idea is illustrated in the top panel of Figure 1. The test response functions describe the expected test score conditional on the trait being measured.  The focal quantity is the gap between the two test response functions, which can be summarized in various ways, typically by taking the expected value of the difference (or its absolute or squared value) over the distribution of the latent trait \cite[e.g.,][]{Chalmers2016, Raju1995}. Regardless of the specific statistic used, the overall idea behind this approach to DTF is to quantify how differential performance at the item-level (i.e., differential item functioning or DIF) affects test scores that aggregate over items.

\begin{figure}[hb] 
\centering
\includegraphics[width = 13cm]{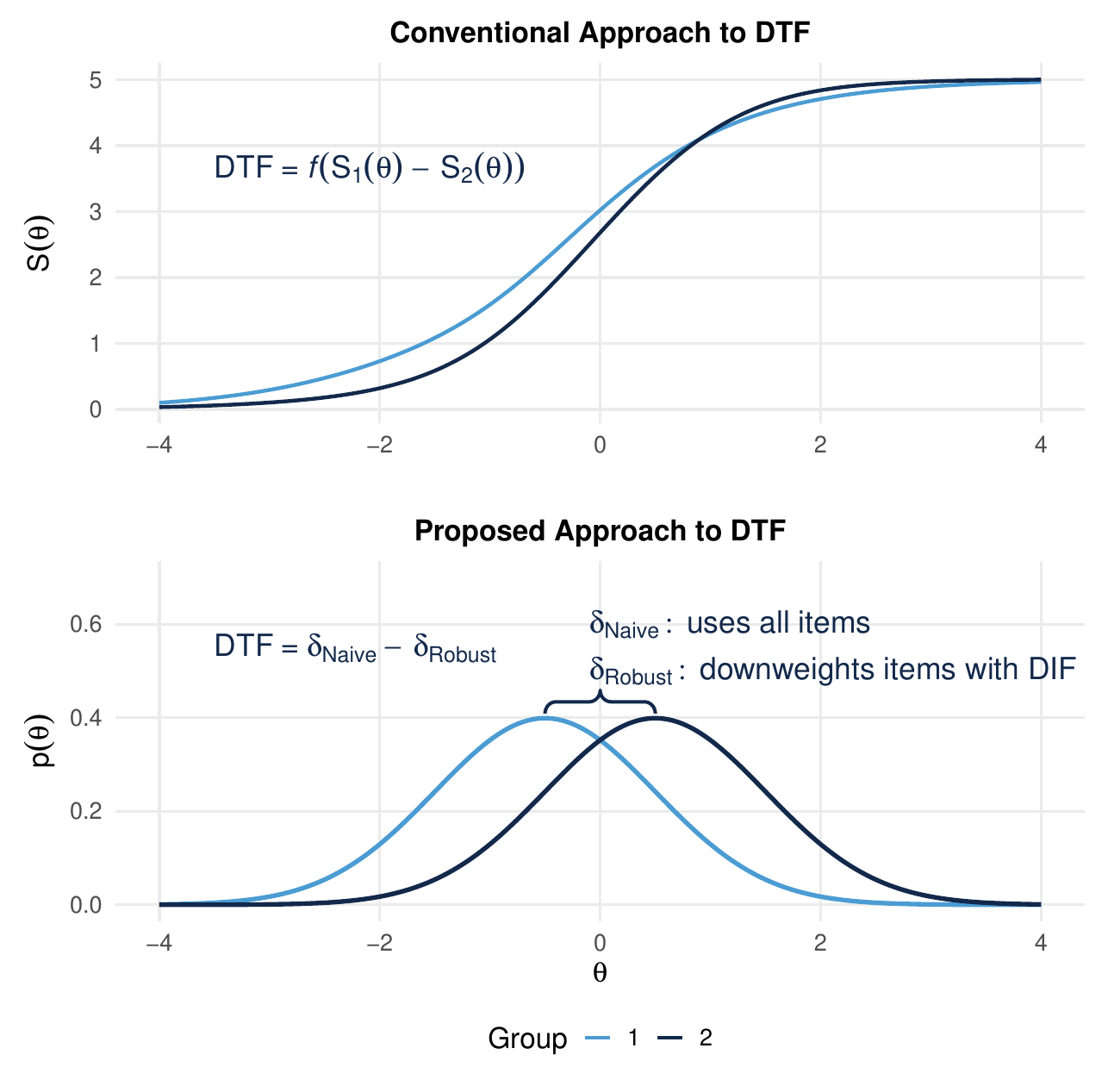}
\caption{Top panel plots the test response function against the latent trait. DTF is defined as a function of the difference between the test response functions in the two groups. Bottom panel shows the distribution of the latent trait in the two groups. The difference between the two distributions is denoted as $\delta$ and DTF is defined as the difference between two ways of defining $\delta$ -- one that uses all assessment items and one that downweights items with DIF.} 
\label{fig:sim1}
\end{figure}

As an alternative, this paper proposes to quantify DTF in terms of the distribution of the latent trait. The overall idea is illustrated in the bottom panel of Figure 1. The difference between the distribution of the latent trait in the two groups has been referred to as impact \citep{Angoff1993}. Impact can be summarized in many ways, and the focus of the present paper is the standardized group-mean difference, denoted $\delta$. As described in more detail below, $\delta$ can be computed by aggregating over assessment items using standard procedures from the literature on test linking and scaling \citep[e.g.,][Chap. 6]{Kolen2014}. Consequently, it is possible to quantify how DIF affects conclusions about impact, analogously to how standard approaches to DTF quantify how DIF affects test scores. We could introduce new terminology to distinguish DTF qua test scores from DTF qua impact, but expanding the notion of DTF to include both approaches emphasizes that they both focus on how DIF aggregates up to the test level. Consequently, this paper uses DTF as an umbrella term for both approaches.   

The proposed approach to DTF involves comparing two estimates of impact. The first estimate naively aggregates over assessment items, while the second is a robust estimate that down-weights items that exhibit DIF. In this paper it is argued that the difference between the naive and robust estimands, denoted $\Delta$, provides a convenient effect size for quantifying the extent to which DIF affects conclusions about impact. A Wald test of the null hypothesis $\Delta = 0$ is also provided.  

As with other approaches to DTF, the methodology developed in this paper can be used as a follow-up to conventional item-by-item DIF analyses. However, the methodology is particularly well suited to robust scaling procedures that are designed to estimate impact in the presence of DIF \citep[e.g.,][]{He2013, He2015, He2020, Halpin2024, Wang2022}. In general, robust scaling procedures do not require conducting item-by-item tests of DIF before estimating impact. Consequently, they avoid the post-selection inference problem \cite[see, e.g., ][chap. 6]{Hastie2015} that arises when selecting anchor items and / or omitting items with DIF prior to evaluating DTF. Being able to infer whether $\Delta = 0$ prior to conducting item-by-item tests of DIF also has practical advantages. For example, in research settings where the overall purpose is to compare groups on the latent trait, testing whether $\Delta = 0$ can indicate whether reported results are robust to DIF without actually having to conduct an item-by-item DIF analysis, which can be time consuming, error prone, and subject to many ``researcher degrees of freedom." 

An additional advantage of robust scaling procedures is that they have been shown to perform well even when relatively large proportions of items exhibit DIF \citep{Halpin2024, Wang2022}. This paper contributes to the literature on robust scaling by presenting a new result on the consistency of the robust estimator developed by Halpin (\citeyear{Halpin2024}). The estimator is shown to be consistent so long as fewer than 1/2 of items exhibit DIF, which is the theoretical upper bound on the number of outliers that can be tolerated by any non-trivial translation equivariant estimator \citep{Huber2009}. While conventional approaches to DIF based on anchor item selection are generally lacking analytical results, extensive simulation studies have shown that they can fail badly when $\geq 1/4$ of items exhibit DIF in the same direction \cite[e.g.,][]{Kopf2015b, Kopf2015a}. More recent approaches to DIF analysis based on penalized-likelihood estimation have been shown to achieve the upper limit of 1/2, but research in this area has not yet addressed consistent estimation of impact \citep{Chen2023}.   

Following discussion of the proposed effect size, the robust scaling procedure, and the Wald test of DTF, this paper presents a simulation study showing how the proposed methodology performs in finite samples. The methodology is then applied to an empirical example from physics education. The procedures are implemented in the R package \texttt{robustDIF} and the replication code for the empirical analysis is provided in the Supplementary Materials. The focus of this paper is the 2PL in two independent groups, with extensions addressed in the conclusion. 

\section{DTF in terms of impact}

Begin by specifying the 2PL model in slope-intercept form:

\begin{equation}\label{2pl}
 \text{logit}(p_{gi}) =   a_{gi} \theta_g + b_{gi},
\end{equation}
where: 
\begin{itemize} 
\item $g = 1, 2$ denotes the reference and comparison groups, respectively. 
\item $i = 1 \dots m$ denotes the assessment items. 
\item $p$ is the probability of endorsing an item. 
\item $a > 0 $ is the item slope (discrimination).  
\item $b \in \mathbb R$ is the item intercept (item threshold or ``easiness''). 
\item $\theta_g = (\theta^*_g- \mu_g)/ \sigma_g$ with $\theta^*_g \sim N(\mu_g, \sigma_g^2)$.  
\end{itemize}

Letting $\bs{\nu} = (a_{11}, b_{11}, \dots a_{1m}, b_{1m}, a_{21}, b_{21}, \dots a_{2m}, b_{2m})^{\top}$ denote the vector of item parameters, it is assumed that maximum likelihood estimates (MLEs) $\hat{\bs{\nu}}$ are available, with asymptotic distribution $\sqrt{n} (\hat{\bs{\nu}} - \bs{\nu}) \overset{d}{\rightarrow} N(0, V(\hat{\bs{\nu}}))$, where $n = n_0 + n_1$ is the total sample size and $n_1 / n_0 \rightarrow (0, \infty)$ \cite[see e.g.,][]{Bock2021}.

As shown in the last bulleted point, the constraints required to identify the model are imposed by standardizing the latent trait in both groups. In the literature on IRT-based scaling, this is referred to as ``separate calibrations'' and can be implemented by either estimating the IRT model separately in each group or using a multi-group approach and scaling the latent trait separately in each group. Standardizing the latent trait separately in each group implies that impact is ignored when estimating the model. However, it is possible to recover group differences on the latent trait in a post-estimation step, which is often referred to as test scaling or linking. 

The remainder of this section outlines a variation on IRT moment-based linking \cite[see][chap. 6]{Kolen2014}. The overall idea is to begin by defining a scaling parameter of interest. In the present case, this will be a standardized group-mean difference on the latent trait. Next we derive \emph{item-level scaling functions} that equate the parameters of each item to the target scaling parameter, under the assumption that the item does not exhibit DIF. Third we aggregate over items to obtain \emph{test-level scaling functions}. Finally we arrive at the proposed effect size for DIF by comparing a naive test-level scaling function that equally weights all items to a robust alternative that downweights items that exhibit DIF. These steps are now explained in more detail. 

Define the scaling parameter of interest as:
\begin{equation} \label{delta_0}
\delta_0 = (\mu_1 - \mu_2) / \sigma.
\end{equation}
The key idea is that the scaling parameter is defined to represent impact on the latent trait. This paper focuses on the case where $\sigma = \bar{\sigma} = \sqrt{(\sigma_1^2 + \sigma_2^2)/2}$ is the pooled standard deviation (SD), so that the magnitude of $\delta_0$ does not depend on the choice of reference group. The overall approach is seen to readily generalize to other choices of scaling parameters \cite[see, e.g., ][]{Halpin2024, Wang2022}. 
  
Next, define the item-level scaling functions:
\begin{equation} \label{delta-i}
\delta_i({\bs{\nu}}) = (b_{i1} - b_{i2})/a_{ip} \quad \text{with} \quad  \bar{a}_i = \sqrt{(a_{i1}^2 +a_{i2}^2) / 2} . 
\end{equation}
In Section \ref{sec:delta} of the Appendix, it is shown that $\delta_i({\bs{\nu}}) = \delta_0$ if and only if item $i$ does not exhibit DIF on either item parameter. This implies that the $\delta_i({\bs{\nu}})$ are all equal to the same constant, $\delta_0$, when no item exhibits DIF. 

Various test-level scaling functions can be obtained by aggregating over assessment items:
\begin{equation} \label{delta}
\delta({\bs{\nu}})= \sum_{i=1}^{m} w_i\, \delta_i({\bs{\nu}}),
\end{equation}
where $w_i  \in [0, 1]$ and $\sum_i w_i = 1$. The assumption that no item exhibits DIF leads to the unweighted mean as an intuitive choice of test-level scaling function.  In the IRT literature, this has been referred to as ``mean scaling" \citep[][chap. 6]{Kolen2014}. The unweighted mean will be written as $\delta_{U}(\bs\nu)$ with weights $w_{Ui} = 1/m$. Because $\delta_{U}(\bs\nu)$ is motivated by the assumption that no item exhibits DIF, it will be referred to as a naive scaling function. 

To address the situation in which some items exhibit DIF, we can re-write Equation \e{delta} as 
\begin{equation}  \label{delta-dif}
\delta({\bs{\nu}}) = \delta_0 + \sum_{i=1}^{m} w_i\, c_i 
\end{equation}
where $c_i = \delta_i({\bs{\nu}}) - \delta_0$ quantifies DIF. From Equation \e{delta-dif}, it is apparent that setting $w_i = 0$ for all items that exhibit DIF (i.e., for which $c_i \neq 0$) will ensure that $\delta({\bs{\nu}})$ remains equivalent to $\delta_0$. In this paper, variations of $\delta({\bs{\nu}})$ that seek to down-weight items with DIF will be referred to as robust scaling functions and denoted using the generic notation $ \delta_R({\bs{\nu}})$ and $w_{Ri}$ for scaling function and weights, respectively. Intuitively, one way to obtain a robust scaling function is to first conduct an item-by-item analysis of DIF and then set $w_{Ri} = 0$ for any items that are inferred to exhibit DIF. The following section presents an alternative approach based on robust scaling. 

The last step is to consider the difference between the naive and robust scaling functions: 
\begin{equation}  \label{Delta}
\Delta (\bs{\nu}) = \delta_{U} (\bs{\nu}) - \delta_{R} (\bs{\nu}) = \sum_{i=1}^{m} \left(1/m - w_{Ri} \right) \delta_i(\bs{\nu}). 
\end{equation} 
The quantity $\Delta (\bs{\nu})$ can be interpreted as an effect size for DTF.  For example, if $\delta_U= 0.5$ and $\delta_R = 0.4$, then $\Delta = 0.1$ means that ignoring DIF would lead us to overestimate the group mean difference on the latent trait by 0.1 SD units. Thus we could conclude that DIF has an effect on conclusions about impact, and that the size of this effect is 0.1 SD units. 

We must be careful about the interpretation of the null hypothesis that $\Delta (\bs{\nu}) = 0$, because this can arise in a number of ways. For example, $\Delta (\bs{\nu}) = 0$ can arise when no items exhibit DIF (i.e., $c_i = 0$ for all items). However, even if $c_i \neq 0$ for some items, $\sum_{i=1}^{m} c_i = 0$ is sufficient for $\delta_U(\bs{\nu}) = \delta_0$. This has been referred to as DIF cancellation \citep{Sireci2013}. If we further assume that $\delta_R(\bs{\nu})$ accurately down weights items with DIF, then it follows that $\delta_R(\bs{\nu}) = \delta_0$ and $\Delta (\bs{\nu}) = 0$ under DIF cancellation. In both of these situations (no DIF and DIF cancellation), $\Delta (\bs{\nu}) = 0$ has the implication that the naive and robust scaling functions are both equivalent to the true impact on the latent trait (i.e., $\delta_{U} (\bs{\nu}) =  \delta_{R} (\bs{\nu}) = \delta_0$). These two situations are summarized under case 1A in Table \ref{tab:Delta}. This case represents the desired interpretation of the null hypothesis. 

However, if we allow for the realistic possibility that the robust scaling function does not down-weight items with perfect accuracy, it is possible to concoct examples where $\delta_{U} (\bs{\nu}) =  \delta_{R} (\bs{\nu}) \neq \delta_0$. For example, using $\delta_i = \{1, 4, 10\}$ and weights $w_i = \{0, 5/6, 1/6\}$, the unweighted mean and weighted mean of the $\delta_i$ are both equal to 5. However, under the assumption that a strict subset of items exhibit DIF, it must be the case that $\delta_0 = \delta_i$ for some value of $\delta_i$, none of which are equal to 5. In this case, it would be misleading to conclude that $\Delta (\bs{\nu}) = 0$ implies that DIF does not affect conclusions about impact -- rather, DIF affects both test-level scaling functions the same way, and neither is equivalent to the true impact on the latent trait. This situation is described under case 1B in Table \ref{tab:Delta}. This case is not the desired interpretation of the null hypothesis. 

\begin{table}[ht] \label{tab:Delta}
\begin{center}
\begin{threeparttable}
\caption{Interpretations of $\Delta$}
\begin{tabular}{lll}
\midrule
Value of $\Delta$ & Cases & Interpretation \\ 
\midrule
Null: $\Delta = 0$ &  1A: $\delta_{U} (\bs{\nu}) =  \delta_{R} (\bs{\nu}) = \delta_0 $ & No DIF / DIF cancellation \\
 & 1B: $\delta_{U} (\bs{\nu}) =  \delta_{R} (\bs{\nu}) \neq \delta_0 $ & DIF affects both $\delta_{U} (\bs{\nu})$ and $ \delta_{R} (\bs{\nu})$ in the same way \\
\midrule
Alternative: $\Delta \neq 0$ 
& 2A: $\delta_{U} (\bs{\nu}) \neq  \delta_{R} (\bs{\nu}) = \delta_0 $ & DIF affects $\delta_{U} (\bs{\nu})$ but not $ \delta_{R} (\bs{\nu})$ \\
& 2B: $\delta_{U} (\bs{\nu}) \neq  \delta_{R} (\bs{\nu}) \neq \delta_0 $ & DIF affects both $\delta_{U}( \bs{\nu})$ and $ \delta_{R} (\bs{\nu})$ \\
& 2C: $\delta_{R} (\bs{\nu}) \neq  \delta_{U} (\bs{\nu}) = \delta_0 $ & DIF affects $\delta_{R} (\bs{\nu})$ but not $ \delta_{U} (\bs{\nu})$ (pathological)\\
 \hline 
\end{tabular}
\end{threeparttable}
\end{center}
\end{table}

To distinguish these two cases, it would be helpful to have theoretical guarantees about the conditions under which $\delta_{R} (\bs{\nu}) = \delta_0$. When these conditions are satisfied, only case 1A remains compatible with $\Delta (\bs{\nu}) = 0$, and the null hypothesis retains its desired interpretation. The robust scaling procedure presented in the following section is seen to provide such a guarantee whenever less than 1/2 of items exhibit DIF. As previously mentioned, analytical results on conventional approaches to DIF  based on anchor item selection are generally lacking, but extensive simulation studies have shown that they can fail badly when  $\geq 1/4$ of items exhibit DIF in the same direction \citep[e.g.,][]{Kopf2015a, Kopf2015b}. 

As shown in Table \ref{tab:Delta}, there are also multiple interpretations of the alternative hypothesis that $\Delta (\bs{\nu}) \neq 0$. Only case 2A is compatible with the assumption that $\delta_{R} (\bs{\nu}) = \delta_0$. Thus, under this assumption, the alternative hypothesis has the interpretation that the naive test-level scaling function is not equivalent to impact on the latent trait. This is also the case for 2B, whereas 2C would indicate pathological behavior of the robust scaling function. 

In summary, this section has proposed to quantify the effects of DIF on conclusions about impact by adapting methods from moment-based linking in IRT. In particular, comparing a naive test-level scaling function that ignores DIF to a robust alternative that down-weights items with DIF has been seen to result in an interpretable effect size for quantifying the effects of DIF when aggregated over items. However, in order for the proposed effect size to have the desired interpretation (i.e., cases 1A and 2A of Table \ref{tab:Delta}), it is necessary to assume that the robust test-level scaling function is equivalent to impact. To this end, the following section describes a robust test-level scaling function that remains consistent for impact so long as fewer than 1/2 of items exhibit DIF. 

\section{Estimating impact using robust scaling} \label{sec:robustDIF}

The ideas in the previous section can be used with various choices of robust scaling functions. This section outlines one choice that is highly robust to DIF \citep{Halpin2024} and presents a new result about its consistency. The construction of the robust scaling function borrows computational techniques from robust estimation of a location parameter \cite[see][chap. 3 \& 4]{Huber2009}. In the present setting, the item-level treatment effects $\delta_i({\bs{\nu}})$ play the role of ``data points'' whose location we wish to ``estimate.'' This analogy is seen to lead to a robust version of Equation \e{delta}.

Begin by writing the weights in Equation \e{delta} in terms of a scalar-valued function $\psi$, to be chosen subsequently: 
\begin{equation} \label{w}
{w}_{Ri} = \frac{\psi(u_i) / u_i}{\sum_{r = 1}^m \psi(u_r) / u_r}.  
\end{equation} 
By convention, the ratio $\psi(u)/u$ is set to 1 when $u = 0$. The argument $u_i = u_i(\bs\nu) = \delta_i(\bs\nu) - \delta$ is a function of both the item parameters $\bs\nu$ and a ``preliminary value'' of the test-level treatment effect, $\delta$. The dependence on $\delta$ leads to iteratively re-weighted least squares (IRLS) as a computational strategy \cite[see][chap. 3]{Huber2009}. 

In the present context, it is desirable to choose $\psi$ to be a so-called \emph{redescending} function \cite[see][section 4.8]{Huber2009}. While many redescending functions are available, Tukey's bi-square function performs well in practice and is convenient for illustrating the main ideas. The bi-square is defined as 
\begin{equation} \label{bsq}
\psi(u) = \left\{\begin{array}{ccc}
u \left(1 - \left( \frac{u}{k} \right)^2\right)^2 & \text{ for   } & |u| \leq k \\		
 0 & \text{ for } &|u| > k \\
\end{array} \right. .
\end{equation}
The overall idea behind the bi-square is to choose the tuning parameter $k$ such that $\psi = 0$ for outliers. This has the effect of setting the weights $w_{Ri} = 0$, so that outliers are down-weighted to zero during estimation. It can also be confirmed that the ratio $\psi(u)/u$ is non-negative and reaches a maximum of 1 when $u = 0$. Thus, in the idealized case that $u_i = 0$ for all items without DIF and $|u_i| > k$ for all items with DIF, the test-level scaling function in Equation \e{delta} becomes the unweighted mean of the items without DIF. 

The performance of the bi-square depends strongly on the choice of the tuning parameter $k$ \citep{Huber1984}. In usual applications, $k$ is chosen to be proportional to the scale (e.g., median absolute deviation) of the data points. In the present context, we can follow a similar logic by choosing $k$ based on the asymptotic distribution of the item-level scaling functions $\hat \delta_i = \delta_i(\hat {\bs{\nu}})$. Under the null hypothesis that item $i$ does not exhibit DIF (i.e., $\delta_i({\bs{\nu}}) = \delta_0$), this distribution can be written as $\sqrt{n}(\hat\delta_i - \delta_0) \overset{d}{\rightarrow} N(0, V_0(\hat\delta_i))$. Letting $\alpha$ denote the desired Type I Error (false positive) rate for down-weighting items with DIF, we may choose item-specific tuning parameters
\begin{equation}\label{k}
k_i = z_{1-\alpha/2} \times \sqrt{V_0(\hat \delta_i) / n }, 
\end{equation} 
where $z_{q} $ denotes the $q$-th quantile of the standard normal distribution. This choice of $k_i$ implies that the weights $w_i$ will be set to zero whenever $\hat \delta_i$ is beyond the $(1 - \alpha) \times 100 \%$ asymptotic confidence interval centered at $\delta$. 

Moving forward, $\delta_{RDIF}({\bs{\nu}})$ will denote the robust scaling function that uses weights $w_{RDIFi}$ defined in Equations \e{w} through \e{k}. The following theorem addresses the consistency of the plug-in estimator $\hat{\delta}_{RDIF} = \delta_{RDIF}(\hat{\bs{\nu}})$. The proof is provided in Section \ref{sec:proof} of the Appendix.

\begin{theorem}
Assume that $d^\star$ is the unique modal (most common) value of $\delta_i(\bs{\nu})$, $i = 1 \dots m$ . Then, as the sample size $n \rightarrow \infty$, the plug-in estimator $\hat \delta_{RDIF}$ defined in Equations \e{delta}, \e{w} - \e{k} converges in probability to $d^\star$. 
\end{theorem} 

The theorem provides a relatively weak condition under which $\hat \delta_{RDIF}$ will remain consistent for the true impact on the latent trait.  In the worst-case scenario that all items that exhibit DIF have item-specific scaling functions equal to the same aberrant value, then $\hat \delta_R$ will remain consistent for $\delta_0$ so long as fewer than 1/2 of items exhibit DIF. Under the assumption that DIF is idiosyncratic -- i.e., $\delta_i(\bs{\nu}) \neq \delta_j(\bs{\nu})$ unless they are both equal to $\delta_0$ --  the theorem implies that $\hat \delta_R$ will remain consistent for $\delta_0$ if just two items do not exhibit DIF. However, it is important to note that the proof of the theorem requires that standard error of the plug-in estimates $\hat\delta_i$ must be substantially smaller than the variation among the true $\delta_i(\bs\nu)$ before the mechanism ensuring consistency ``kicks in." Consequently, the performance of the estimator with realistic sample sizes is an important consideration. This is addressed in the simulation studies reported below. 

Theorem 1 implies that the DTF effect size $\Delta_{RDIF} = \delta_U - \delta_{RDIF}$ has the desired interpretation (cases 1A and 2A in Table \ref{tab:Delta}) whenever fewer than 1/2 of items exhibit DIF. This leaves open the question what to do when 1/2 or more items exhibit DIF, which is topic of future research that is revisited in the conclusion. 

\section{A Wald test of DTF} \label{sec:wald}

A Wald test of $\Delta (\bs{\nu}) = 0$ is available via the asymptotic distribution of the plug-in estimator $\Delta (\hat{\bs{\nu}})$, which can be obtained by applying the Delta method \cite[e.g.,][chap. 3]{Vaart1998}. The Delta method gives
\begin{equation}\label{test} 
\sqrt{n} (\hat \Delta  - \Delta (\bs{\nu})) \overset{d}{\rightarrow} N(0, V(\hat \Delta)) 
\end{equation}  
where $V(\hat\Delta) = [\partial \Delta (\bs{\nu})]^{T} \, V(\hat{\bs{\nu}})  [\partial \Delta (\bs{\nu})]$ and $\partial \Delta (\bs{\nu})$ is the column vector of partial derivatives of $\Delta (\bs{\nu})$. The derivation of   $\partial \Delta (\bs{\nu})$ when using $\delta_{RDIF}$ as the robust scaling function is provided in section \ref{sec:comp} of the Appendix. The relevant result is  
\begin{equation}\label{var.Delta} 
V(\hat \Delta_{RDIF}) = \sum_{i = 1}^m \sum_{j = 1}^m (1/m - v_i)(1/m - v_j) \, \text{COV}(\hat{\delta}_i, \hat{\delta}_j) \qquad \text{with} \qquad v_i = \frac{\psi'(u_i) / V_0(\hat{\delta}_i) }{\sum_{r = 1}^{m}\psi'(u_r) / V_0(\hat{\delta}_r)}
\end{equation}  
It is important to note that $\psi'(u) = d\psi/du $ can take on negative values with redescending loss functions. This is problematic because it means that the denominator of $v_i$ can approach zero, leading $V(\hat \Delta_{RDIF})$ to diverge to infinity.  
This concern is especially relevant when the loss function is tuned to aggressively flag outliers \citep[][\S 4.8]{Huber2009}, which is the case here. To address this issue, negative values of $\psi'(u)$ are set to zero when computing $v_i$. This approach stabilizes of $V(\hat \Delta_{RDIF})$ at the expense of reducing the effective number of items used for computation. As shown in the simulation studies reported below, this approach achieves close-to-nominal rejection rates under the null hypothesis while retaining the desirable robustness properties of $\hat\delta_{RDIF}$ when many items exhibit DIF. 

Before moving on, let us note how the proposed Wald test be implemented as a follow-up to item-by-item tests of DIF. The overall idea is that, rather than using the robust scaling function $\hat\delta_{RDIF}$, we will first manually test for DIF in each item and then downweight to zero any items that are inferred exhibit DIF to obtain an alternative robust scaling function, $\hat\delta_{R}$. Letting $m_0$ denote the number of items inferred not to exhibit DIF, the DIF-corrected scaling function be computed using weights $w_{Ri} = 0$ for items that exhibit DIF and $w_{Ri} = 1/m_0$ otherwise. The variance of the effect size $\hat \Delta = \hat\delta_U - \hat\delta_R$ can be computed by substituting $w_{Ri} = v_i$ into Equation \e{var.Delta}.  As previously noted, this approach ignores uncertainty in the item-by-item tests of DIF by treating the $w_{Ri}$ as known constants. Consequently, the resulting expression for the $V(\Delta)$ will be overly optimistic (i.e., too small), which is a known problem in post-selection inference \cite[][chap. 6]{Hastie2015}. One advantage of the robust scaling approach is that it avoids this problem.  
 
\section{Simulation Study}

\subsection{Design}
The main goal of the simulation study is to complement Theorem 1 by describing the finite sample performance of $\hat\delta_{RDIF}$ and its associated Wald test of DTF under varying sample sizes and varying degrees of DIF. Data were generated using the parameters given in Table \ref{tab:sim}. One focal factor was sample size, $n_g = \{200, 350, 500\}$, which was held equal over groups. The second focal factor was the proportion of items with DIF, which ranged from 0 to $1/2$. Items with DIF were randomly selected in each replication. The item-level effects were held constant at $c_i = \delta_i - \delta_0 = 0.5$, corresponding to the ``worst-case scenario'' described under Theorem 1.  DIF was induced by adding the appropriate constant to the item intercepts in the second group when generating the data. Thus, the simulation is limited to consideration of uniform DIF. 

\begin{table}[ht] \label{tab:sim}
\begin{center}
\begin{threeparttable}
\caption{Summary of Simulation Design.}
\begin{tabular}{ll}
 \hline
Design factor & Value \\ 
\hline
Respondents per group (focal) &  $n_1 = n_2 = \{200, 350, 500\}$ \\
Proportion of items with DIF (focal) & 0 to 1/2, items chosen randomly\\ 
Number of items & $m = 20$ \\ 
Item slopes & $a \sim U(0.5, 2)$ and $a_{1} = a_{2} = a$ \\
Item intercepts (without DIF) &  $b^* \sim U(-1.5, 1.5)$, $b_{1} = b_{2} = a b^*$ \\
Latent trait & $\theta_1 \sim N(0, 1)$, $\theta_2 \sim N(\mu, 1)$, $\mu = U(-.25, 25)$\\
Replications per condition & 500 \\
   \hline 
\end{tabular}
\begin{tablenotes}
\item \emph{Note}: Items with DIF were generated by adding a constant to the item intercepts such that $\delta_i - \delta_0 = 0.5$ for each item. 
\end{tablenotes}
\end{threeparttable}
\end{center}
\end{table}

The simulations also considered two comparison methods in which items with DIF were treated as known when computing the proposed Wald test of DTF. As outlined above, this amounts to computing the test-level scaling function in Equation \e{delta} as the unweighted mean of the items that are assumed not to exhibit DIF. One comparison method used the true DIF pattern in the data generating model. This method serves as a check on correctness of the proposed methodology. The second comparison method used the likelihood-ratio test (LRT) test of DIF on item intercepts \citep{Thissen1993}. This methods serves to illustrate the proposed methodology with one widely-used DIF procedure. The LRT test was implemented using two-stage purification and refinement \cite[]{Holland1993}. In the first stage, all items other than the tested item were used as anchors. In the second stage, any items inferred to exhibit DIF in the first stage were omitted from the anchor and all items were retested. An item was omitted from the anchor if its LRT was significant at the 5\% level. In the second stage, false discovery rate was controlled at 5\% using the Benjamini-Hochberg procedure \citep{Benjamini1995}. 

IRT model estimation and the LRT procedure were implemented using the \text{mirt} \citep{Chalmers2012}. The RDIF procedure and Wald tests of DTF were implemented using the \texttt{robustDIF} package in \texttt{R}. Replication code is provided in the Supplementary Materials. 
  
 \subsection{Results}

\begin{figure}[h] 
\centering
\includegraphics[width = 16.6cm]{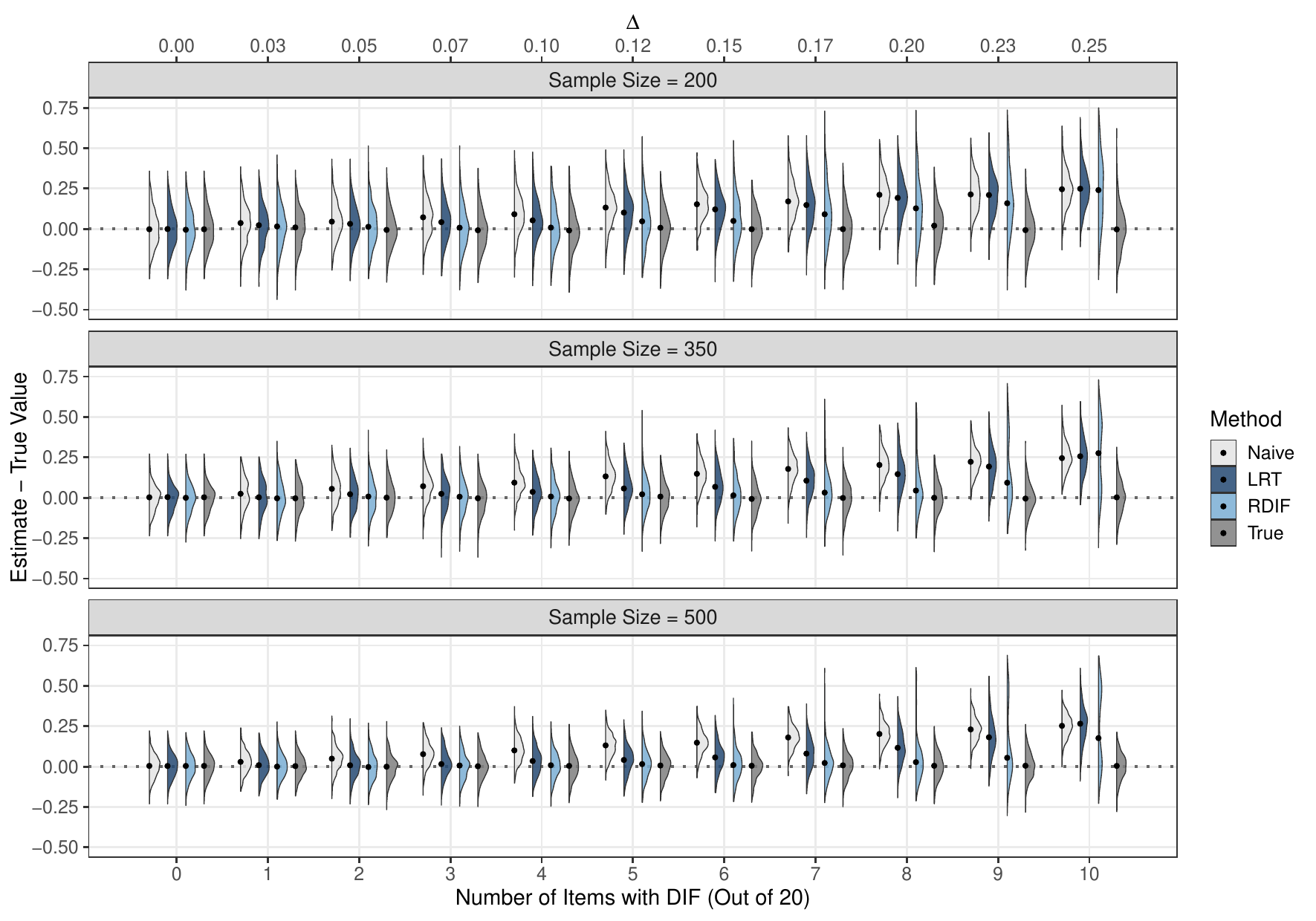}
\caption{Bias and variance of the different scaling functions. Naive denotes the unweighted mean of the item-level scaling functions. LRT denotes the scaling functions computed by down-weighting items flagged using the likelihood ratio test. RDIF denotes the robust scaling procedure described in Section \ref{sec:robustDIF} of this paper. True denotes the scaling functions compute by down-weighting items with DIF in data generating process (i.e., without inference). The values of $\Delta$ in the top axis describe the true degree of bias in the naive estimate. Tabular results are presented in section \ref{sec:tables} of the Appendix.} 
\label{fig:bias}
\end{figure}

Figure \ref{fig:bias} shows the bias and variance of the four scaling functions. Focussing first on the bottom panel (the largest sample size), the following observations can be made. In the condition where no items exhibit DIF, all estimates were unbiased and had similar variance. Also note that, when no items exhibit DIF, the naive and true estimates are equivalent, and both are equivalent to the LRT estimate in the absence of false positives. Thus, in the conditions without DIF, we are essentially comparing only the naive and RDIF estimates. 

As the number of items with DIF increases, the naive estimates becoming increasingly biased but their variance remains approximately constant, because all items are used to compute these estimates. By contrast, the True estimates remain unbiased and their variance increases, because they omit any items with DIF in the data generating process. 

The relation between the LRT and RDIF can be interpreted in terms of a bias-variance trade-off: the RDIF estimates are generally less biased but the LRT estimates are generally more precise. This pattern becomes more pronounced as the number of items with DIF increases. When the proportion of items with DIF reaches 1/2, the distribution of the RDIF estimates becomes markedly bi-modal. This behavior is aligned with Theorem 1, which implies that the RDIF estimate will be inconsistent in this condition, with local minima at the two modal values $\delta_i(\bs{\nu}) = \{\delta_0, \delta_0 + 0.5\}$. These local minima are the modes of the distribution the shown figure. 

The same patterns are apparent but less pronounced at the smaller sample sizes. At the smallest sample size in particular, the relative unbiasedness of the robust DIF estimates compared to the LRT estimates is quite diminished. Overall, the comparison between the LRT and robust DIF estimates indicates that the robust DIF procedure is advantageous when the proportion of items with DIF is at least $1/4$ and the sample size is at least $350$ per group. 
 
Figure \ref{fig:power} shows the error rates for the Wald tests of DTF using a nominal Type I Error rate of 5\%. In terms of Type I Error, both the LRT- and the RDIF-based tests maintained nominal level quite well at all sample sizes. For the LRT-based test, this was mainly attributable to the FDR correction. Using more lenient criteria for flagging items with DIF (e.g., no FDR correction, information-criteria), resulted in over-rejection using the LRT. For the RDIF-based test, these results indicate that the computational approximation used for the asymptotic variance performed quite well in the absence of DIF, even at the smallest sample size. Results for the True DIF pattern are omitted when no items exhibit DIF, because in this case the distribution of the test statistic is degenerate. 

In terms of power, the overall conclusion is that the LRT-based test was more powerful than the RDIF-based test when $\leq 6/20$ items exhibited DIF, but the pattern was reversed for larger proportions of DIF. At the smallest sample size neither method was well-powered, but power did reach acceptable levels for the larger sample sizes. It is interesting that the RDIF-based test maintained acceptable power even when 1/2 of items exhibited DIF. In this case, the distribution of the RDIF scaling function becomes bimodal (see Figure \ref{fig:bias}), but in this simulation both modes were equidistant from the expected value of the naive estimate. This resulted in reasonable power, although this is likely an artifact of this specific study design. The test based on the True DIF pattern shows that proposed procedure is quite powerful in the idealized setting where the items with DIF are known a priori, and suggests that there is room to improve the test by improving inference about DIF. 

In terms of practical recommendations, the LRT-based method with item-level false positive control can be recommended when fewer than 6/20 (or approximately 1/4) of items exhibit DIF, the RDIF is preferred when larger proportions of items exhibit DIF, and both methods required a sample size of about $350$ per group to attain power of $0.8$ using a nominal 5\% false positive rate. 

\begin{figure}[h] 
\centering
\includegraphics[width = 15cm]{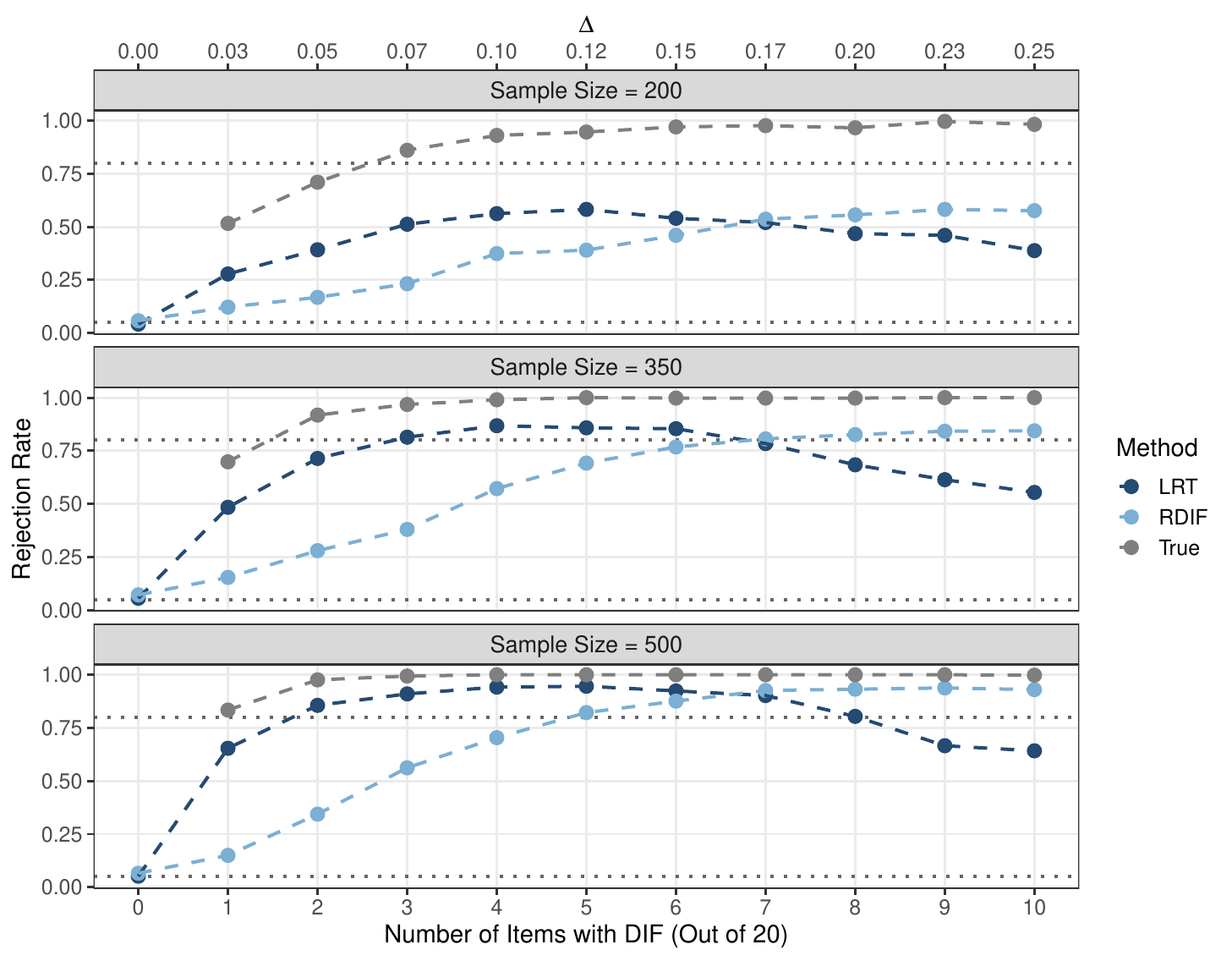}
\caption{Rejection rates for Wald test of the null hypothesis that $\Delta = 0$ using a nominal rejection rate of $\alpha = .05$. LRT denotes the test statistic computed as a follow-up to the likelihood ratio test of DIF. RDIF denotes the robust scaling procedure described in Section \ref{sec:robustDIF} of this paper. True denotes the test statistic computed by flagging items with DIF based on the data generated process (i.e. without inference). The true effect sizes are labelled in the top axis. Tabular results are presented in section \ref{sec:tables} of the Appendix. }
\label{fig:power}
\end{figure}

\subsection{Summary}
The main goal of the simulation study was to complement Theorem 1 by describing the finite sample performance of the RDIF scaling function and associated Wald test of DTF under varying sample sizes and varying degrees of DIF. The theorem states that the RDIF scaling function will be consistent for the true scaling parameter so long as fewer than 1/2 of items exhibit DIF, but the result requires that the item-level scaling functions have small standard errors relative to the variability among their true values. The simulation indicated that a sample size of 350 was sufficient for the RDF estimate to remain relatively unbiased when fewer than 1/2 of items exhibit DIF equivalent to 0.5 SD units on the latent trait. It can be anticipated that larger sample sizes would be required for smaller DIF effects. Compared to the LRT, the relative unbiasedness of the RDIF scaling function was most notable when the proportion of items with DIF exceeded 1/4. Similarly,  the RDIF-based Wald test has superior power when the proportion of items with DIF exceeded 6/20 (approximately 1/4), although a sample size of 350 per group was required to attain acceptable levels of power. 

It can be concluded that the RDIF procedure is to be preferred when sample sizes are relatively large ($\approx$ 350 per group) and the proportion of items with DIF may exceed 1/4. It should be cautioned that this simulation did not address unequal sample sizes or varying magnitudes of DIF, due to the large number of possible combinations of these factors. In applied settings, simulation-based power analyses reflecting the characteristics of specific studies can be conducted using the \texttt{robustDIF} software. 

\section{Empirical example} 
The example examines gender-based DTF on the Force Concept Inventory \citep[FCI;][]{Hestenes1992}. The FCI is widely used in college-level physics education. It is intended to test knowledge of basic concepts in Newtonian physics using real-world examples and minimizing mathematical formalities. It has been  documented to exhibit gender-based DIF \citep{Traxler2018}, and there have been concerns about the extent to which gender gap on the overall test reflects differences in physics knowledge as opposed to other sources of variation \citep{Madsen2013}. This example illustrates the proposed Wald test of DTF using a sample of $n_1 = 360$ females and $n_2 = 371$ males who were assessed at the beginning of the Fall 2013 semester. The data are not publicly available due to Family Educational Rights and Privacy Act (FERPA) restrictions in the United States. 

Table \ref{tab:dtf} summarizes the results of the Wald test of DIF. It can be seen that the LRT- and RDIF-based approaches both lead to the conclusion that the unweighted mean of the test items (i.e., the naive estimate) over-estimates the gender gap in physics knowledge (males - females) by about 0.1 SD units on the latent trait. However, after correcting for DIF, both the RDIF and LRT-based estimates indicate that a substantial gender gap of 0.88 SD remains. Table \ref{tab:dif} shows the item-level tests of DIF based on the RDIF procedure \citep[see][]{Halpin2024} and the LRT. The item-level tests lead to similar conclusions overall. The RDIF item-level effects $\hat\delta_i - \hat\delta_R$ indicate whether the item-level gender gap was larger (positive) or smaller (negative) than the test-level gender gap based on the RDIF estimate. The majority of items with DIF were biased towards males.  

The overall conclusions is that, while item-level DIF favoring males led to upward bias of the estimated gender gap by 0.1 SD units on the latent, a sizable gender gap remained after correcting for DIF (.88 SD units on the latent trait).   

\begin{table}[ht] \label{tab:dtf}
\begin{center}
\begin{threeparttable}
\caption{Tests of Differential Test Functioning}
\begin{tabular}{rrrrrrrrr}
  \toprule
 & Naive Est. & Naive SE & DIF Est. & DIF SE & $\hat\Delta$ & $SE(\hat\Delta)$ & $z$ & $p(>|z|)$ \\ 
  \midrule
  RDIF & 0.997 & 0.087 & 0.884 & 0.093 & 0.112 & 0.045 & 2.477 & 0.013 \\ 
  LRT & 0.997 & 0.087 & 0.888 & 0.089 & 0.109 & 0.031 & 3.574 & 0.000 \\ 
   \bottomrule
\end{tabular}
\begin{tablenotes}
\item \emph{Note}: The naive estimate and standard error (SE) are based on the unweighted mean of the item-level scaling functions. The DIF estimate and SE correct for DIF, using either the RDIF approach outlined in this paper or by omitting items with DIF identified by the likelihood ratio test (LRT) of item intercepts. The LRT used two-stage purification and refinement with the Benjamini-Hochberg correction. Positive estimates indicate that male students performed better than female students. 
\end{tablenotes}
\end{threeparttable}
\end{center}
\end{table}

\begin{table}[ht] \label{tab:dif}
\begin{center}
\begin{threeparttable}
\caption{Item-level tests of DIF}
\begin{tabular}{rlrrrrrr}
  \toprule
 & & \multicolumn{4}{c}{RDIF Wald Test} & \multicolumn{2}{c}{LRT} \\
\cmidrule(lr){3-6} \cmidrule(lr){7-8}
 & Items & $\hat\delta_i - \hat\delta_R$ & $SE(\hat\delta_i - \hat\delta_R)$ & $z$ & $p(>|z|)$ & $\chi^2$ & $p(>\chi^2)$ \\ 
  \midrule
01 & Dropping unequal metal balls & 0.364 & 0.174 & 2.091 & \bf{0.037} & 5.356 & \bf{0.021} \\ 
02 & Rolling balls off table & -0.051 & 0.175 & -0.294 & 0.769 & 0.131 & 0.717 \\ 
03 & Why objects accelerate downward & 0.182 & 0.184 & 0.991 & 0.322 & 0.805 & 0.370 \\ 
04 & Truck–car collision forces & -0.119 & 0.183 & -0.653 & 0.514 & 1.529 & 0.216 \\ 
05 & Forces inside curved channel & -0.421 & 0.208 & -2.026 &\bf{0.043} & 2.405 & 0.121 \\ 
06 & Ball exiting curved track & 0.825 & 0.218 & 3.785 & \bf{0.000} & 10.166 & \bf{0.001} \\ 
07 & String breaks during swing & -0.029 & 0.156 & -0.189 & 0.850 & 0.402 & 0.526 \\ 
08 & Kicked moving hockey puck & 0.112 & 0.160 & 0.702 & 0.483 & 0.190 & 0.663 \\ 
09 & Speed after sideways kick & -0.045 & 0.285 & -0.157 & 0.875 & 0.065 & 0.799 \\ 
10 & Speed after impulsive kick & 0.064 & 0.107 & 0.600 & 0.548 & 0.087 & 0.768 \\ 
11 & Forces on sliding puck & -0.400 & 0.176 & -2.266 & \bf{0.023} & 10.864 & \bf{0.001} \\ 
12 & Cannonball off a cliff & 0.629 & 0.194 & 3.248 & \bf{0.001} & 8.991 & \bf{0.003} \\ 
13 & Thrown ball in midair & -0.173 & 0.139 & -1.240 & 0.215 & 5.605 & \bf{0.018} \\ 
14 & Object dropped from airplane & 0.802 & 0.185 & 4.344 & \bf{0.000} & 14.133 & \bf{0.000} \\ 
15 & Pushing truck while accelerating & 0.052 & 0.227 & 0.231 & 0.818 & 0.098 & 0.754 \\ 
16 & Pushing truck at constant speed & 0.032 & 0.152 & 0.213 & 0.832 & 0.155 & 0.694 \\ 
17 & Elevator moving at constant speed & -0.169 & 0.241 & -0.703 & 0.482 & 1.913 & 0.167 \\ 
18 & Forces on swinging child & -0.505 & 0.182 & -2.768 & \bf{0.006} & 10.857 & \bf{0.001} \\ 
19 & Comparing block speeds & -0.081 & 0.180 & -0.451 & 0.652 & 0.998 & 0.318 \\ 
20 & Comparing block accelerations & 0.138 & 0.139 & 0.993 & 0.321 & 0.249 & 0.618 \\ 
21 & Spaceship with sideways thrust & 0.662 & 0.272 & 2.438 & \bf{0.015} & 6.197 & \bf{0.013} \\ 
22 & Spaceship speed under thrust & 0.303 & 0.184 & 1.651 & 0.099 & 2.433 & 0.119 \\ 
23 & Spaceship after engine cutoff & 0.769 & 0.158 & 4.878 & \bf{0.000} & 26.341 & \bf{0.000} \\ 
24 & Post-thrust spaceship speed & 0.366 & 0.131 & 2.790 & \bf{0.005} & 6.652 & \bf{0.010} \\ 
25 & Pushing box at constant speed & -0.141 & 0.200 & -0.706 & 0.480 & 0.598 & 0.439 \\ 
26 & Doubling push on box & -0.133 & 0.285 & -0.464 & 0.642 & 3.981 & \bf{0.046} \\ 
27 & Stopping applied force & 0.778 & 0.281 & 2.769 & \bf{0.006} & 5.998 & \bf{0.014} \\ 
28 & Students pushing off chairs & 0.168 & 0.123 & 1.369 & 0.171 & 0.030 & 0.863 \\ 
29 & Forces on resting chair & -0.448 & 0.200 & -2.239 & \bf{0.025} & 5.909 & \bf{0.015} \\ 
30 & Forces on flying tennis ball & -0.169 & 0.186 & -0.904 & 0.366 & 1.582 & 0.208 \\ 
  \bottomrule
\end{tabular}
\begin{tablenotes}
\item \emph{Note}: LRT denotes the likelihood ratio test of the item intercepts. Bolded p-values are less than .05 (without correcting for false discovery rate). Positive (negative) values of $\hat\delta_i - \hat\delta_R$ indicate that the item-level gender gap (male - female) was larger (smaller) than the robust test-level gender gap. \end{tablenotes}
\end{threeparttable}
\end{center}
\end{table}

\section{Conclusion}
This paper has proposed a conceptualization of DTF that focusses on impact (i.e., the distribution of the latent trait) whereas previous approaches have focussed on test scores. Both approaches share the main idea of quantifying the consequences of DIF when aggregating over items. The proposed approach adapts well-known methods for moment-based scaling in IRT \cite[e.g., ][chap. 6]{Kolen2014} to define item-level scaling functions that are equivalent to a standardized group-mean difference on the latent trait in the absence of DIF. DTF was defined by comparing two methods of aggregating these item-level scaling functions, one that naively averages over all items and one that downweights items with DIF. The resulting effect size describes the extent to which ignoring DIF would affect conclusions about how groups differ on the latent trait. 

In practice, the interpretation of the proposed effect size depends on accurately downweighting items with DIF. Conventional approaches for DIF analysis based on anchor item selection can fail badly when the proportion of items with DIF exceeds 1/4  \cite[e.g.,][]{Kopf2015b, Kopf2015a}. For higher proportions of DIF, performance can be improved using robust scaling procedures that automatically downweight items when estimating impact. In particular, Theorem 1 of this paper showed that the RDIF scaling procedure \citep{Halpin2024} remains consistent for impact on the latent trait whenever fewer than 1/2 of items exhibit DIF. Simulation studies supported the conclusion that the RDIF estimator of impact can be substantially less biased than approaches based on anchor item selection when the proportion of items with DIF exceeds 1/4 and the sample size is at least 350 per group. 

The proposed effect size leads to a convenient Wald test of DTF.  Simulations showed that the Wald test based on the RDIF estimator was less powerful than conventional approaches when the proportion of items with DIF $\lesssim 1/4$, but was more powerful at higher proportions. Both approaches required sample sizes of $\approx 350$ to achieve acceptable power, although the simulations indicated that power at lower sample sizes could be substantially improved via more accurate inference about DIF. 

It is important to note that the simulation study reported in this paper was limited in several ways (equal sample sizes per group, uniform DIF, size and direction of DIF). Rather than generalizing based on this single study, it is recommended to conduct simulation-based power analyses tailored to the details of a given application. Implementation is available via the \texttt{robustDIF} package in \texttt{R} and illustrated in the Supplemental Materials.  

Other limitations of this work that can be addressed by future research include extensions of the methodology to a wider range of models, analytical results on the robustness of the RDIF procedure when more than 1/2 of items  exhibit empirically plausible patterns of DIF, investigation of other robust scaling methods with improved efficiency, and DTF methodology that can accommodate multiple and possibly dependent groups (e.g., longitudinal invariance).

\section{Appendix}

\subsection{Item-level scaling functions }\label{sec:delta} 
This section derives the item-level scaling function in Equation \e{delta-i} and shows that it is  equal to the target scaling parameter $\delta_0$ if and only if the item does not exhibit DIF on either item parameter. 
 
Models that can be represented using Equation \e{2pl} are identified only up to an affine transformation of the latent trait \cite[see, e.g.,][]{Bechger2015, vanderLinden2016}. Consequently, we may write 
\begin{equation} \label{scale0}
a_{ig} \theta_g + b_{ig} = a^*_{ig} \theta^*_g + b^*_{ig} 
\end{equation}
in which $\theta_g = (\theta^*_g- \mu_g)/ \sigma_g$ and $g = 1, 2$. In this set-up, the parameters with without asterisks are scaled using standardized $\theta$ and those with asterisks are scaled using $\theta^*_g  \sim N(\mu_g, \sigma_g)$ where the scaling parameters $\mu_g$ and $\sigma_g$ are unknown constants. For the remainder of this section I omit the subscript for items to simplify notation. 

Algebraic manipulation of Equation \e{scale0} leads to 
\begin{equation} \label{scale1}
\mu_1 - \mu_2 = \frac{a^*_{2} (b_{1} - b_{1}^*) - a^*_{1} (b_{2} - b_{2}^*)}{a^*_{1}  a^*_{2}} 
\end{equation}
and
\begin{equation} \label{scale2}
\bar \sigma = \sqrt{\frac{\sigma_1^2 + \sigma_2^2}{2}} =  \sqrt{\frac{(a_{1}  a^*_{2})^2 + (a^*_{1}  a_{2})^2}{2 a^*_{1}  a^*_{2}}} 
\end{equation}
so that 
\begin{equation} \label{scale3}
\delta_0 \equiv  \frac{\mu_1 - \mu_2}{\bar \sigma} = \frac{a^*_{2} (b_{1} - b_{1}^*) - a^*_{1} (b_{2} - b_{2}^*)}{\sqrt{((a_{1}  a^*_{2})^2 + (a^*_{1}  a_{2})^2)/2}}. 
\end{equation}
The assumption that the item does not exhibit DIF means that the item parameters are equal across groups prior to standardization of the latent trait -- i.e., $a^*_{1} = a^*_{2} = a^*_{}$ and $b_{1}^* = b_{2}^* = b_{}^*$. Applying these substitutions to Equation \e{scale3} gives 
\begin{equation} \label{scale4}
\delta_0 = \frac{b_{1} - b_{2}}{\sqrt{(a_{1}^2 + a_{2}^2)/2}} . 
\end{equation}
The right hand side of this expression is the item-level scaling function introduced in Equation \e{delta-i} of the main paper. Consequently, Equation \e{scale4} shows sufficiency: if we assume the item does not exhibit DIF, then the item-level scaling function is equal to $\delta_0$.

To show necessity, we start with the assumption that item-level scaling function is equal to $\delta_0$ and show that this implies no DIF. It is convenient to re-parameterize using $a^*_{1} = a^*$, $a^*_{2} = A a^*$,  $b^*_{1} = b^*$, and $b^*_{2} = b^* + B$ and then show that $A = 1$ and $B = 0$. The derivation requires several steps but uses only basic algebra. 

Using Equation \e{scale3} with the new parameterization, the assumption is
\begin{equation} \label{scale5}
 \frac{b_{1} - b_{2}}{\sqrt{a_{1}^2 + a_{2}^2}} = \frac{A a^*(b_{1} - b^*) - a^* (b_{2} - b^* - B)}{\sqrt{(A a_{1}  a^*)^2 + (a_{2}a^*)^2}} = \frac{K}{\sqrt{A^2 a_{1}^2 + a_{2}^2}} 
\end{equation}
%
%
%
where the last equality factors out $a^*$ from the numerator and denominator and uses $K = A (b_{1} - b^*) - b_{2} + b^* + B$.  For the case where $b_{1} = b_{2} = b$, Equation \e{scale5} implies $K = 0$, which can be re-written as 
\begin{equation} \label{scale6}
 (A-1)(b - b^*) + B = 0
\end{equation}
This equality holds for all values of $b - b^*$ only if $A = 1$ and $B = 0$, as required.  For the case where $b_{1} \neq b_{2}$, we cross-multiply in Equation \e{scale5}, square both sides, and collect terms to get
%
%
%
\begin{equation} \label{scale8}
(K^2 - A^2 (b_{1} -b_{2})^2)a_{1}^2 +  (K^2 - (b_{1} -b_{2})^2) a_{2}^2  =  0. 
\end{equation}
In order to hold for Equation \e{scale8} to hold for all $a_{1}^2$ and $a_{2}^2$, we must have 
\begin{equation} \label{scale6}
K^2 - A^2 (b_{1} -b_{2})^2 = 0 \quad \text{and} \quad K^2 - (b_{1} -b_{2})^2 =  0.
\end{equation}
Since $b_1 \neq b_2$, the equality $A^2 (b_{1} -b_{2})^2 = (b_{1} -b_{2})^2$ shows that $A = 1$. Substituting this value into $K$ we get $K = b_1 - b_2 + B$ so that Equation \e{scale5} holds only if $B = 0 $.

\subsection{Proof of Theorem 1} \label{sec:proof}

The set-up for the proof is adapted from Huber's (\citeyear{Huber1984}) discussion of the finite sample breakdown of M-estimators of location. This approach makes use of  $\rho = \int \psi(u) du$, where $\psi$ was used to define the weights $w_{Ri}$ in Equation \e{w}. Defining
\begin{equation} \label{R}
R(\delta; \bs{\nu}) = \sum_{i=1}^m \rho(\delta_i(\bs{\nu})- \delta),
\end{equation} 
the RDIF scaling function can be written (omitting subscripts) as 
\begin{equation} \label{R2}
\delta(\bs{\nu}) = \underset{\delta \in \mathbb{R}}{\text{arg min}} \; R(\delta; \bs{\nu}) \end{equation} 
This loss function representation implies the weighted mean representation in Equation \e{delta} \cite[see][\S 3.2]{Huber2009} but is more suitable to proving the theorem. 

To address the case of redescending M-estimators, it is assumed that $\rho$ satisfies the following assumptions: 
\begin{itemize} 
\item[A1] $\rho: \mathbb R \rightarrow [-1, 0]$ is continuous.
\item[A2] $\rho(0) = -1$ is the unique minimum.
\item[A3] There exists a value of $k > 0$ such that $\rho(u) = 0$ for $|u| > k$. 
\end{itemize} 
These assumptions include the (appropriately rescaled) bi-square function defined in Equation \e{bsq} as well as other commonly used redescending M-estimators \cite[see][section 4.8]{Huber2009}.  

The proof appeals to Theorem 5.7 of van der Vaart (\citeyear{Vaart1998}), which states two conditions that are together sufficient to ensure the consistency of M-estimators. One condition is that the population loss function must have a unique global minimum, say $\delta^\star$. The second condition is that the sample loss function must converge uniformly  in probability to the population loss function. Together these conditions imply that the minimizing argument of the sample loss function converges in probability to $\delta^\star$. The following two lemmas serve to establish the population loss function for $\delta(\bs{\nu})$ and uniform convergence for $\delta(\hat{\bs{\nu}})$, respectively. 

\begin{lemma}  Let $\delta_i(\bs{\nu})$, $i = 1, \dots m$, denote a finite collection of data points and let $S(\bs{\nu})$ denote the set of unique values of $\delta_i(\bs{\nu})$. For each $s \in S(\bs{\nu})$ let $m_s$ denote the number of data points such that $\delta_i(\bs{\nu}) = s$ and let $\epsilon$ be the smallest distance between any two values $s_i, s_j \in S(\bs{\nu})$. If we choose $k < \epsilon / 2$, then $R(s; \bs{\nu}) = -m_s$ is a local minimum of $R(\delta; \bs{\nu})$ for each $s \in S(\bs{\nu})$, where $R(\delta; \bs{\nu})$ is defined by Equation \e{R} and assumptions A1-A3.  
\end{lemma} 

\textbf{Proof of Lemma 1.} A2 implies that $\rho(\delta_i(\bs{\nu})- s) = -1$ for $\delta_i(\bs{\nu}) = s$, while A3 and the choice of $k$ imply that $\rho(\delta_j(\bs{\nu})- s) = 0$ for $x_j \neq s$. Thus, $R(s) = -m_s$ is the minimum of $R$ (by A2) on the interval $[s - \epsilon/2, s + \epsilon/2]$ (by A3 and choice of $k$). $\square$ 

An immediate corollary of Lemma 1 is that the population loss function will have a unique global minimum when there is a unique modal (most frequent) value of $\delta_i(\bs{\nu})$. This is the condition stated in Theorem 1. Another important corollary of the Lemma 1 is that global minimum of $R(\delta; \bs{\nu})$ must lie in the finite set $S(\bs{\nu})$, which is used in the following lemma. 

\begin{lemma}. Let $R(\delta; \hat{\bs{\nu}})$ be the loss function in Equation \e{R} computed using the MLEs $\hat{\bs{\nu}}$ and tuning parameter $k_i^* = \max\{k_i, \epsilon/2\}$ with $k_i$ defined in Equation \e{k} and $\epsilon$ defined in Lemma 1. Then $R(\delta; \hat{\bs{\nu}})$ converges uniformly in probability on the finite set $S(\bs{\nu})$ to the population loss function $R(\delta; \bs{\nu})$ defined in Lemma 1. 
\end{lemma} 

\textbf{Proof of Lemma 2.} By continuity of $\rho$ (A1) and the continuous mapping theorem applied to the MLEs $\hat\nu$, we have point-wise convergence for each fixed value of $\delta$:
\[ 
R(\delta; \hat{\bs{\nu}}) - R(\delta; \bs{\nu})| \overset{p}{\rightarrow} 0
\] 
Additionally, from Equation \e{k}, it is apparent that $k_i \rightarrow 0$ as $n \rightarrow \infty$. Thus, $k^*_i \rightarrow \epsilon/2$ and the tuning parameter of $R(\delta; \bs{\nu})$ deterministically satisfies the condition stated in Lemma 1. Consequently, the global minimum of $R(\delta; \bs{\nu})$ lies in the finite set $S(\bs{\nu})$. Point-wise convergence then implies
\[ 
\underset{\delta \in S(\bs{\nu})}{\max} |R(\delta; \hat{\bs{\nu}}) - R(\delta; \bs{\nu})|\overset{p}{\rightarrow} 0,
\] 
which establishes uniform convergence on the finite set $S(\bs{\nu})$.  

The proof is not entirely satisfying because there no mechanism describing how one transitions from optimization on  $\delta \in \mathbb{R}$ in Equation \e{R2} to $\delta \in S(\bs{\nu})$ in Lemma 2. However, Lemma 1 establishes $\delta_R(\bs{\nu}) \in S(\bs{\nu})$, implying that it is indeed sufficient to consider optimization of Equation \e{R2} on $S(\bs{\nu})$. A second limitation has to do with the definition of the modified tuning parameter $k_i^*$, which requires $\epsilon$ and hence also depends on knowledge of $S(\bs{\nu})$. Better proofs may be available. 

Together with the condition stated in Theorem 1, the two lemmas imply that $\delta(\hat {\bs{\nu}})$ satisfies conditions stated in Theorem 5.7 of van der Vaart (\citeyear{Vaart1998}). 

\subsection{Computational details} \label{sec:comp} 

The asymptotic distribution of the RDIF estimator described in the main text was derived in Halpin (\citeyear{Halpin2024}) using the Delta method \cite[e.g.,][chap. 3]{Vaart1998}. The relevant result is
\begin{equation} \label{var-delta}
V(\delta_{RDIF}(\hat{\bs{\nu}})) = \sum_{i = 1}^m [v_i\nabla \delta_i(\bs{\nu})]^T  V(\hat{\bs{\nu}} ) [v_i\nabla \delta_i(\bs{\nu})]  
\end{equation}
with $v_i$ defined in Equation \e{var.Delta} of this paper. Consequently, we require only the derivatives of the item-level scaling functions $ \delta_i(\bs{\nu})$, which are are listed below. 
\begin{align} \notag
\frac{\partial}{\partial a_{i1}} \delta_{i}  &= -\delta_i  \frac{a_{i1}}{a_{i1}^2 + a_{i2}^2}   \\\notag
\frac{\partial}{\partial b_{i1}} \delta_{i} &=  \frac{1}{\sqrt{(a_{i1}^2 + a_{i2}^2)/2}} \\\notag
\frac{\partial}{\partial a_{i2}} \delta_{i} &=   -\delta_i  \frac{a_{i2}}{a_{i1}^2 + a_{i2}^2}  \\
\frac{\partial}{\partial b_{i2}} \delta_{i} &=   -\frac{1}{\sqrt{(a_{i1}^2 + a_{i2}^2)/2}}
\end{align}

Note that Halpin (2024) used the (unstated) assumption that the $\delta_i$ were independent to further simplify Equation \e{var-delta} and considered different scaling functions than those used in this paper. 

 \subsection{Tabular Results from Simulation Study} \label{sec:tables}

\begin{table}[h!] \label{tab:bias}
\begin{center}
\begin{threeparttable}
\caption{Bias and Standard Errors of Test-Level Scaling Functions}
\begin{tabular}{rrrrrrr}
\midrule
  Sample Size & N.DIF & Naive & LRT & RDIF & True \\ 
 \midrule
      &    0 &   0.002 (0.117)  & 0.003 (0.117) & 0.002  (0.124) &  0.002 (0.117)  \\ 
      &    1 &   0.033 (0.119)  & 0.023 (0.121) & 0.013  (0.130) &  0.009 (0.120) \\ 
      &    2 &   0.051 (0.114)  & 0.033 (0.117) & 0.017  (0.126) &  0.002 (0.116) \\ 
      &    3 &   0.071 (0.121)  & 0.042 (0.127) & 0.012  (0.132) & -0.006 (0.124) \\ 
      &    4 &   0.094 (0.122)  & 0.059 (0.128) & 0.019  (0.138) & -0.008 (0.125) \\ 
  200 &    5 &   0.131 (0.117)  & 0.093 (0.127) & 0.051  (0.144) &  0.006 (0.122) \\ 
      &    6 &   0.153 (0.104)  & 0.117 (0.116) & 0.060  (0.145) &  0.002 (0.112) \\ 
      &    7 &   0.178 (0.121)  & 0.149 (0.135) & 0.103  (0.176) &  0.003 (0.127) \\ 
      &    8 &   0.212 (0.118)  & 0.191 (0.132) & 0.139  (0.187) &  0.015 (0.126) \\ 
      &    9 &   0.216 (0.116)  & 0.205 (0.123) & 0.178  (0.199) & -0.008 (0.123) \\ 
      &   10 &   0.251 (0.117)  & 0.253 (0.128) & 0.242  (0.218) &  0.003 (0.136) \\ 
      &    0 &  -0.001 (0.083)  &--0.001(0.083) & -0.002 (0.090) &  -0.001 (0.083) \\ 
      &    1 &   0.027 (0.096)  &-0.011 (0.098) & 0.002  (0.102) &  0.002 (0.097) \\ 
      &    2 &   0.053 (0.093)  & 0.022 (0.096) & 0.008  (0.102) &  0.002 (0.094) \\ 
      &    3 &   0.067 (0.092)  & 0.021 (0.099) & 0.005  (0.102) & -0.006 (0.095) \\ 
      &    4 &   0.099 (0.088)  & 0.038 (0.096) & 0.009  (0.104) & -0.002 (0.091) \\ 
  350 &    5 &   0.132 (0.089)  & 0.062 (0.100) & 0.024  (0.106) &  0.006 (0.093) \\ 
      &    6 &   0.148 (0.090)  & 0.071 (0.105) & 0.021  (0.115) & -0.005 (0.094) \\ 
      &    7 &   0.177 (0.089)  & 0.104 (0.103) & 0.047  (0.141) & -0.000 (0.095) \\ 
      &    8 &   0.205 (0.090)  & 0.148 (0.115) & 0.083  (0.170) &  0.004 (0.098) \\ 
      &    9 &   0.224 (0.086)  & 0.192 (0.112) & 0.150  (0.209) & -0.004 (0.091) \\ 
      &   10 &   0.253 (0.090)  & 0.258 (0.117) & 0.260  (0.225) &  0.001 (0.100) \\ 
      &    0 &   0.003 (0.072)  & 0.003 (0.073) & 0.001  (0.077) & 0.003 (0.072) \\ 
      &    1 &   0.030 (0.071)  & 0.010 (0.072) & 0.002  (0.074) &  0.005 (0.070) \\ 
      &    2 &   0.047 (0.076)  & 0.008 (0.078) & -0.000 (0.083) & -0.003 (0.077) \\ 
      &    3 &   0.079 (0.074)  & 0.022 (0.080) & 0.005  (0.082) &  0.004 (0.077) \\ 
      &    4 &   0.105 (0.079)  & 0.032 (0.083) & 0.010  (0.083) &  0.005 (0.080) \\ 
  500 &    5 &   0.130 (0.073)  & 0.041 (0.081) & 0.016  (0.089) &  0.004 (0.074) \\ 
      &    6 &   0.149 (0.075)  & 0.057 (0.087) & 0.014  (0.095) &  0.003 (0.078) \\ 
      &    7 &   0.182 (0.071)  & 0.080 (0.094) & 0.028  (0.104) &  0.006 (0.077) \\ 
      &    8 &   0.204 (0.074)  & 0.121 (0.102) & 0.052  (0.150) &  0.005 (0.080) \\ 
      &    9 &   0.228 (0.075)  & 0.186 (0.118) & 0.133  (0.217) &  0.006 (0.082) \\ 
      &   10 &   0.250 (0.073)  & 0.259 (0.115) & 0.234  (0.241) &  0.001 (0.081) \\
 \hline
\end{tabular}
\begin{tablenotes}
\item \emph{Note}: Table reports bias and standard errors (SEs, in parentheses) based on 500 simulation replications per condition. Bias and SEs were computed by subtracting the estimate from the data-generating value. Sample size is per group and N.DIF denotes the number of items with DIF (out of 20). Naive denotes the unweighted mean of the item-level scaling functions. LRT denotes the scaling functions computed by down-weighting items flagged by the likelihood ratio test. RDIF denotes the robust scaling procedure described in this paper. True denotes the scaling functions computed by down-weighting items with DIF in data generating process (i.e., without inference).  
\end{tablenotes}
\end{threeparttable}
\end{center}
\end{table}

\begin{table}[h!] \label{tab:power}
\begin{center}
\begin{threeparttable}
\caption{Rejection Rates}
\begin{tabular}{rrrrr}
\midrule
  Sample Size & N.DIF & LRT & RDIF & True \\ 
\midrule 
  &    0 & 0.042 & 0.058 & NA  \\ 
  &    1 & 0.278 & 0.122 & 0.516 \\ 
  &    2 & 0.392 & 0.168 & 0.710 \\ 
  &    3 & 0.512 & 0.232 & 0.860 \\ 
  &    4 & 0.562 & 0.374 & 0.930 \\ 
 200  &    5 & 0.582 & 0.390 & 0.946 \\ 
  &    6 & 0.540 & 0.460 & 0.970 \\ 
  &    7 & 0.520 & 0.536 & 0.976 \\ 
  &    8 & 0.468 & 0.556 & 0.966 \\ 
  &    9 & 0.460 & 0.582 & 0.996 \\ 
  &   10 & 0.388 & 0.576 & 0.982 \\ 
  &    0 & 0.056 & 0.067 &  NA \\ 
  &    1 & 0.484 & 0.154 & 0.698 \\ 
  &    2 & 0.714 & 0.280 & 0.918 \\ 
 &    3 & 0.814 & 0.380 & 0.968 \\ 
  &    4 & 0.868 & 0.572 & 0.990 \\ 
350   &    5 & 0.858 & 0.692 & 1.000 \\ 
  &    6 & 0.854 & 0.768 & 0.998 \\ 
  &    7 & 0.784 & 0.806 & 0.998 \\ 
  &    8 & 0.684 & 0.826 & 0.998 \\ 
  &    9 & 0.614 & 0.842 & 1.000 \\ 
  &   10 & 0.554 & 0.844 & 1.000 \\ 
  &    0 & 0.052 & 0.066 &  NA \\ 
  &    1 & 0.654 & 0.150 & 0.834 \\ 
   &    2 & 0.856 & 0.344 & 0.976 \\ 
  &    3 & 0.910 & 0.562 & 0.994 \\ 
  &    4 & 0.942 & 0.704 & 1.000 \\ 
500  &    5 & 0.946 & 0.822 & 1.000 \\ 
  &    6 & 0.924 & 0.876 & 1.000 \\ 
  &    7 & 0.902 & 0.926 & 1.000 \\ 
  &    8 & 0.804 & 0.932 & 1.000 \\ 
  &    9 & 0.666 & 0.938 & 1.000 \\ 
  &   10 & 0.642 & 0.930 & 0.998 \\ 
   \hline
\end{tabular}
\begin{tablenotes}
\item \emph{Note}: Table reports bias and standard errors (SEs, in parentheses) based on 500 simulation replications per condition. Bias and SEs were computed by subtracting the estimate from the data-generating value. Sample size is per group and N.DIF denotes the number of items with DIF (out of 20). Naive denotes the unweighted mean of the item-level scaling functions. LRT denotes the scaling functions computed by down-weighting items flagged by the likelihood ratio test. RDIF denotes the robust scaling procedure described in this paper. True denotes the scaling functions computed by down-weighting items with DIF in data generating process (i.e., without inference).  
\end{tablenotes}
\end{threeparttable}
\end{center}
\end{table}

\newpage
\FloatBarrier
\bibliographystyle{apalike}


\bibliography{/Users/peterhalpin/Dropbox/Academic/Manuscripts/Bibliography/library} 

\end{document}